\documentclass[twocolumn,floatfix,longbibliography]{revtex4-2}
\usepackage{graphics,epsfig,amsfonts,amssymb,amsmath,xcolor,ulem}
\newcommand{\beginsupplement}{%
        \setcounter{table}{0}
        \renewcommand{\thetable}{S\arabic{table}}%
        \setcounter{figure}{0}
        \renewcommand{\thefigure}{S\arabic{figure}}%
          \setcounter{equation}{0}
        \renewcommand{\theequation}{S\arabic{equation}}
        }

\begin{document}
\title{Correlated states in magic angle twisted bilayer graphene under the optical conductivity scrutiny}
\author{Mar\'ia J. Calder\'on$^1$*}
\email{calderon@icmm.csic.es}
\author{Elena Bascones$^1$*}
\email{leni.bascones@icmm.csic.es}
\affiliation{$^1$Instituto de Ciencia de Materiales de Madrid (ICMM). Consejo Superior de Investigaciones Cient\'ificas (CSIC), Sor Juana In\'es de la Cruz 3, 28049 Madrid (Spain).}
\date{\today}
\begin{abstract}
{\bf Moir\'e systems displaying flat bands have emerged as novel platforms to study correlated electron phenomena. Insulating and superconducting states appear upon doping magic angle twisted bilayer graphene (TBG), and there is evidence of correlation induced effects at the charge neutrality point (CNP) which could originate from spontaneous symmetry breaking. Our theoretical calculations show how optical conductivity measurements can distinguish different symmetry breaking states, 
and reveal the nature of the correlated states.
In the specific case of nematic order, which breaks the discrete rotational symmetry of the lattice, we find that the Dirac cones are displaced, not only in momentum space but also in energy, inducing finite 
Drude weight at the CNP. 
We also show that the sign of the Drude weight anisotropy induced by a nematic order depends on the degree of lattice relaxation, the doping and the nature of the symmetry breaking. }
\end{abstract}
\maketitle
\section{INTRODUCTION}

Instabilities in correlated materials arise when the interaction energy overcomes the kinetic energy gain. Hence, materials whose bands have a very small bandwidth (flat bands) are prone to show correlation effects. The band structure of some moir\'e systems can be tuned to display very narrow bands. A moir\'e pattern appears when there is a small mismatch between two overlapping grids, as it happens in twisted bilayer graphene with one layer rotated by a relative angle $\theta$ with respect to the other. Very weakly dispersing bands  are found in the so-called magic angle twisted bilayer graphene (TBG) with $\theta \sim 1.1^o$ around charge neutrality~\cite{BistritzerPNAS2011}. A plethora of insulating and superconducting states appear as the system is doped~\cite{CaoNat2018_2,CaoNat2018_1,YankowitzScience2019,LuNat2019,StepanovArXiv2019,SaitoArXiv2019} and band widening has been observed in STM experiments when the chemical potential $\mu$ lies within the flat bands, in particular, in undoped systems (namely, at the charge neutrality point CNP)~\cite{KerelskyNat2019,ChoiNatPhys2019,JiangNat2019,XieNat2019}.  Some observations are sample dependent and are affected by whether the TBG is aligned to the underlying Boron Nitride (h-BN) as in-plane twofold rotation symmetry may then be broken~\cite{SharpeScience2019}.  Correlations have to be included to explain both the insulating states which appear at integer fillings  and the doping dependence of the band widening~\cite{PoPRX2018,PizarroJPhysComm2019,YuanPRB2018,KennesPRX2018}. The nature of these states is highly debated. 
An important issue, yet also unknown, is whether the correlated states affect only the flat bands or also involve higher energy bands, split by a few tens of meV gaps from the flat bands. 

\begin{figure*}
\includegraphics[clip,width=\textwidth]{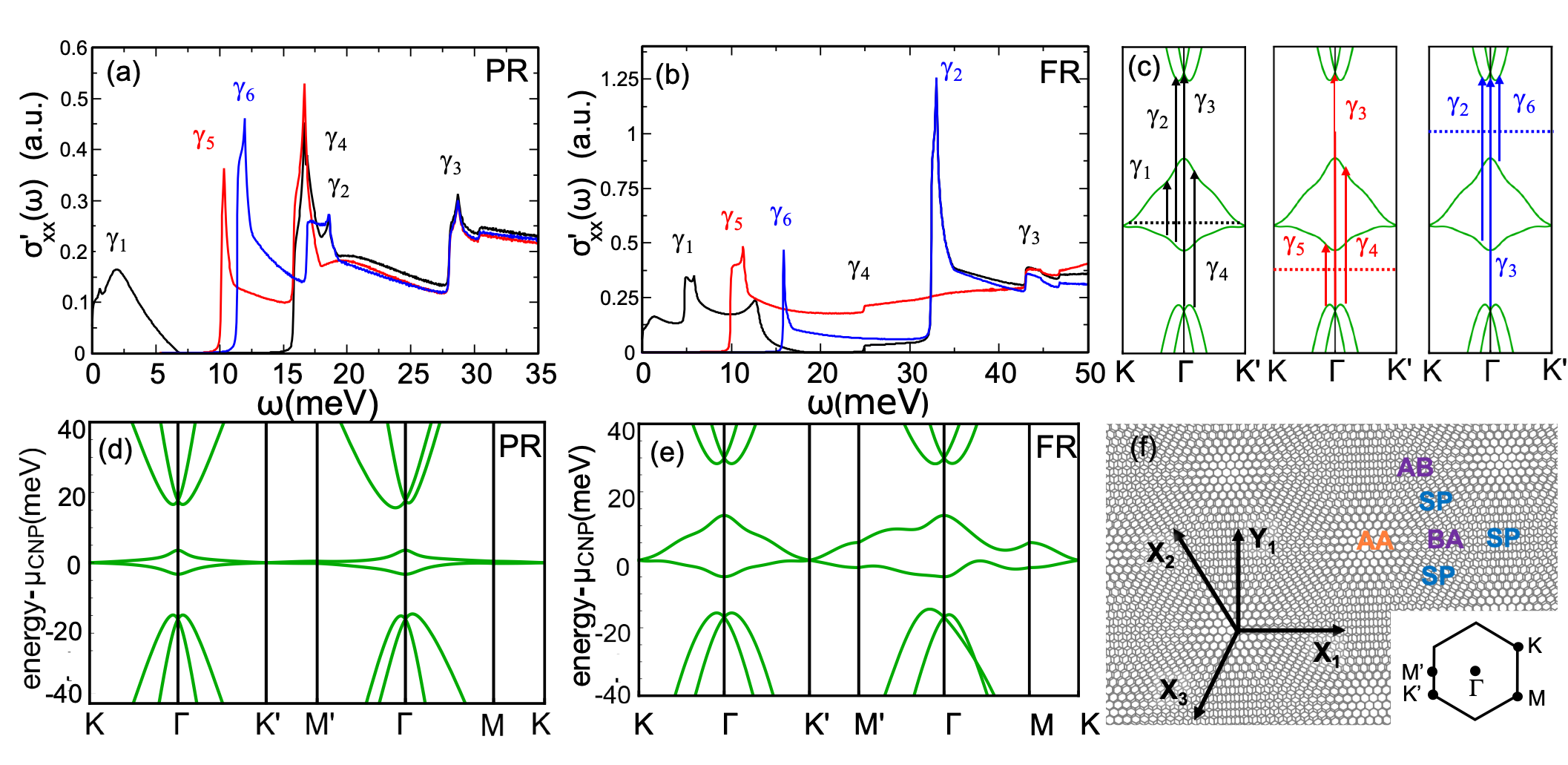}
\caption{{\bf Optical conductivity and band structure of TBG in the non-correlated state.} (a) and (b) Optical conductivity for the non-interacting bands at the CNP (black), for empty flat bands (red) and for full flat bands (blue), respectively, for the partially relaxed (PR) and to the fully relaxed (FR) model.  The peaks in the spectra are labelled with $\gamma_i$ as the transition in (c) from which they originate.  (c) Inter-band transitions that occur at CNP, empty flat bands and filled flat bands. The dashed lines signal the chemical potential $\mu$. (d) Flat bands corresponding to the PR model and (e)  FR model. $\mu_{\rm CNP}$ is the chemical potential at the charge neutrality point. (f) Top view of the stacking order in TBG highlighting the AA, AB and saddle point (SP) regions and sketch of the C$_3$ related $X_i$ directions and of $Y_1$ and Brillouin zone of TBG with the corresponding symmetry points.}
\label{fig:Fig1} 
\end{figure*}

In the absence of symmetry breaking, the upper and lower flat bands touch at Dirac points and are four fold degenerate, stemming from the spin and the graphene valley degrees of freedom. For each valley the bands show M$_{\rm 2y}$, C$_3$ and C$_2$T symmetry~\cite{ZouPRB2018}. Here M$_{\rm 2y}$ is a mirror symmetry which flips the y coordinate producing a layer exchanging two-fold rotation in 3D,  C$_3$ and C$_2$ rotations are referred to an axis perpendicular to the TBG and T stands for time reversal. Signatures of nematicity (C$_3$ symmetry breaking~\cite{FernandesAnn2019}) have been found in measurements of the superconducting upper critical field in doped samples and observed in STM experiments when the chemical potential lies in the flat bands~\cite{CaoArXiv2020,ChoiNatPhys2019,JiangNat2019}. The latter happens at the same dopings at which STM measurements detect the band widening, namely an increased separation of the density of states (DOS) peaks ~\cite{KerelskyNat2019,ChoiNatPhys2019,JiangNat2019,XieNat2019}. 
These effects disappear if the flat bands are completely filled or empty and therefore they cannot be a simple consequence of strain.  On the other hand, activated behavior at the CNP appears in some transport experiments~\cite{LuNat2019,StepanovArXiv2019}, reflecting a gap opening at the Dirac points. Whether this latter effect has an electronic origin or is due to coupling to the substrate has not been clarified. Nevertheless, no alignment with the h-BN has been detected in the experiments. 

Here we study the effect of symmetry breaking states on the optical response of two tight binding models proposed for TBG with different degrees of relaxation.  
We show that the optical conductivity corresponding to C$_2$T and C$_3$ symmetry breaking states display specific signatures that help identify these correlated states. In particular, optical conductivity is sensitive to the gapping of the Dirac points and to the breaking of discrete rotational symmetries.  Unexpectedly, we find that a nematic order can induce Fermi pockets, and hence a finite Drude weight at the CNP. The dependence of the optical conductivity on doping gives information about whether the rotational symmetry breaking arises from the lattice or from the electronic degree of freedom and can clarify if the higher energy bands participate in the correlated state. 
Our calculations show a 
strong sensitivity of the nematic response to the still not well understood lattice relaxation, as well as to the nature of the nematic order. This sensitivity can be traced back to the flat bands structure and impacts dramatically the sign of the Drude weight and the dc conductivity anisotropies of doped TBG.

\section*{RESULTS}

Optical conductivity experiments allow to probe the excitation spectrum of materials~\cite{BasovRevModPhys2011}. Specifically, they probe the $q=0$ inter-band transitions between occupied and empty electronic states, giving accurate measurements of the direct excitation gap in insulators. If a material undergoes a phase transition, its band-structure is modified and the inter-band transitions are sensitive to these changes, including a possible gap opening at low energies. In particular, the  changes in  the band structure responsible for the  modification of the DOS observed in STM when the  flat bands are partially filled, with respect to the one in fully empty or filled flat bands, can be addressed by optical conductivity measurements  in the far-infrared regime. In the following, we present results for the optical conductivity  calculated in linear response as given by Kubo formula, following the derivation for multiorbital systems in Ref.~\cite{ValenzuelaPRB2013}, see Methods and Supplementary Equation 20.  

\subsection*{Optical conductivity in non-correlated states}
\begin{figure*}
\includegraphics[clip,width=\textwidth]{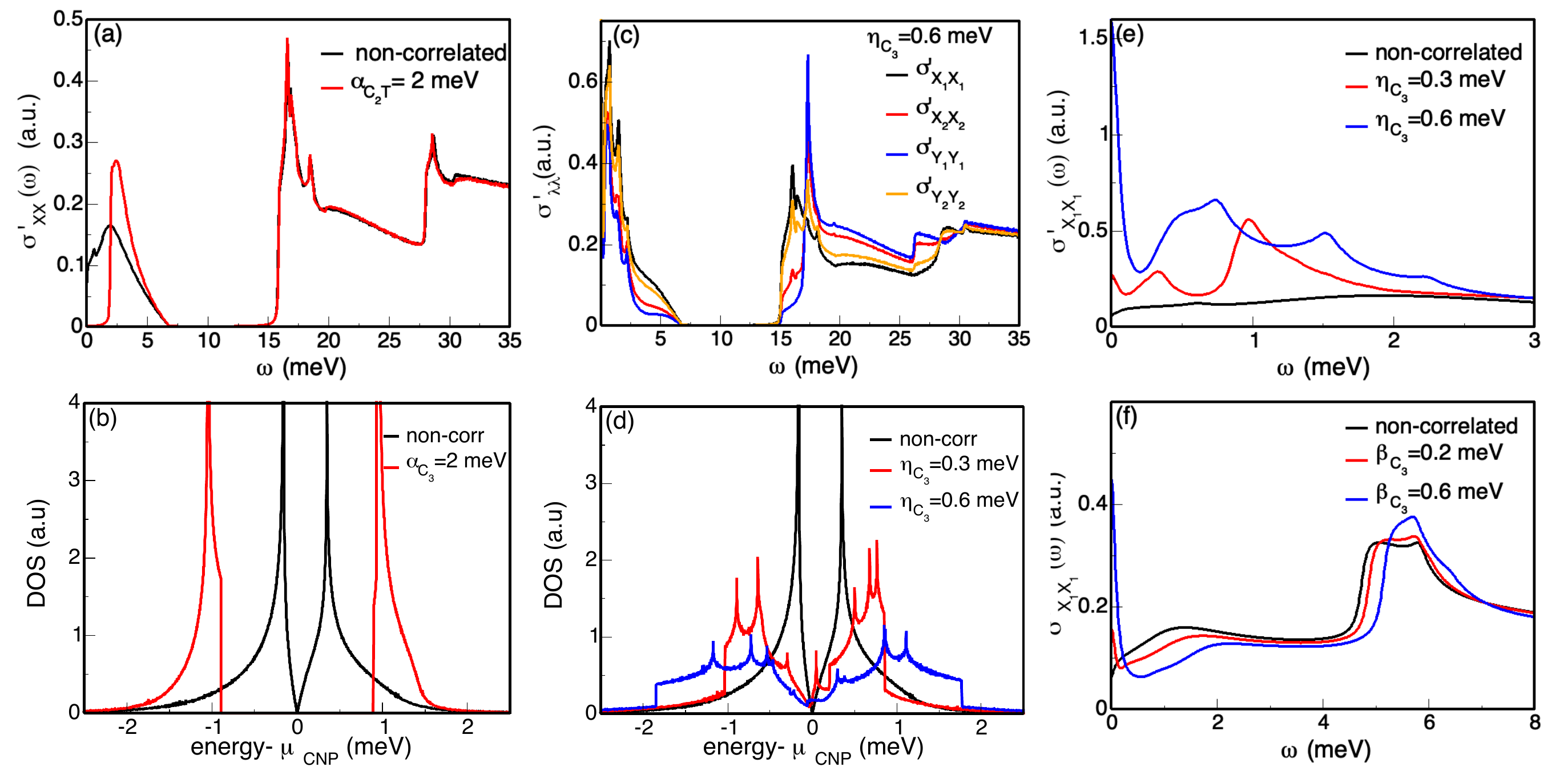}
\caption{{\bf Optical conductivity and density of states of undoped TBG with different symmetry breaking orders.} Comparison of the optical conductivity (a) and the low energy density of states (b) of the partially relaxed (PR) model at the CNP in a C$_2$T symmetry breaking state $\alpha$ and in the non-correlated state. (c) and (d) Same as (a) and (b) but for the nematic  state $\eta$. The optical conductivity in (c) is plotted along four different directions, see Fig.~\ref{fig:Fig1}~(f). In the non-correlated and  the C$_2$T symmetry breaking states, the spectra along these four directions are equal. (e) and (f) Emergence of a Drude peak at charge neutrality. The optical conductivity in the $X_1$ direction at CNP at low frequencies for different values of the nematic order parameter is shown in (e) for the PR model in the state $\eta$ and in (f) for the fully relaxed (FR) model in the state $\beta$. In all the panels $\alpha_{\rm C_2T}$, $\eta_{\rm C_3}$ and $\beta_{\rm C_3}$ control the magnitude of the corresponding order parameter. }
\label{fig:Fig2} 
\end{figure*}

Fig.~\ref{fig:Fig1} (a) and (b) show the optical conductivity of two model systems for TBG, in the absence of correlations, for different values of the electronic filling. 
We focus on $\sigma'_{\lambda\lambda}(\omega)$, the real part of the longitudinal electrical conductivity $\sigma_{\lambda\lambda}(\omega)$ which relates the electric current $J^\lambda$ generated in linear response in direction $\lambda$ and the homogeneous electric field $E^{\lambda}$ applied in the same direction $J^\alpha=\sigma_{\lambda \lambda} (\omega)E^\lambda$. 
In metals, $\sigma'_{\lambda \lambda} (\omega)$ shows a peak at zero frequency, the Drude peak,  related to the d.c. conductivity. At higher energies there is contribution from the inter-band transitions with energy $\hbar \omega$ and zero momentum $q=0$, weighted by matrix elements that depend on the electronic momentum $k$ and on the orbital composition of the bands. Following Pauli exclusion principle, the allowed transitions depend on doping. The arrows in Fig.~\ref{fig:Fig1}~(c) indicate the optical transitions at CNP (black), with empty flat bands (red) and with full flat bands (blue). The optical transitions are labelled $\gamma_m$ with $m$ ranging from $1$ to $6$.

The non-interacting band structures of the two model systems, a fully relaxed (FR)  TBG with $\theta_{\rm FR}\sim  0.9^o$  and  a partially relaxed (PR) TBG with $\theta_{\rm PR}\sim 1.05^o$, are displayed, for a single valley, in Fig.~\ref{fig:Fig1} (d) and (e) and in the Supplementary Information~\cite{PoPRB2019,CarrPRR2019}.
In both models  the chemical potential of the undoped system $\mu_{\rm CNP}$  touches the Dirac points, located at the K and K' points of the Brillouin zone. Besides the different bandwidths, a clear contrast between both models is the reduced particle-hole symmetry of the FR model. 

\begin{figure*}
\includegraphics[clip,width=\textwidth]{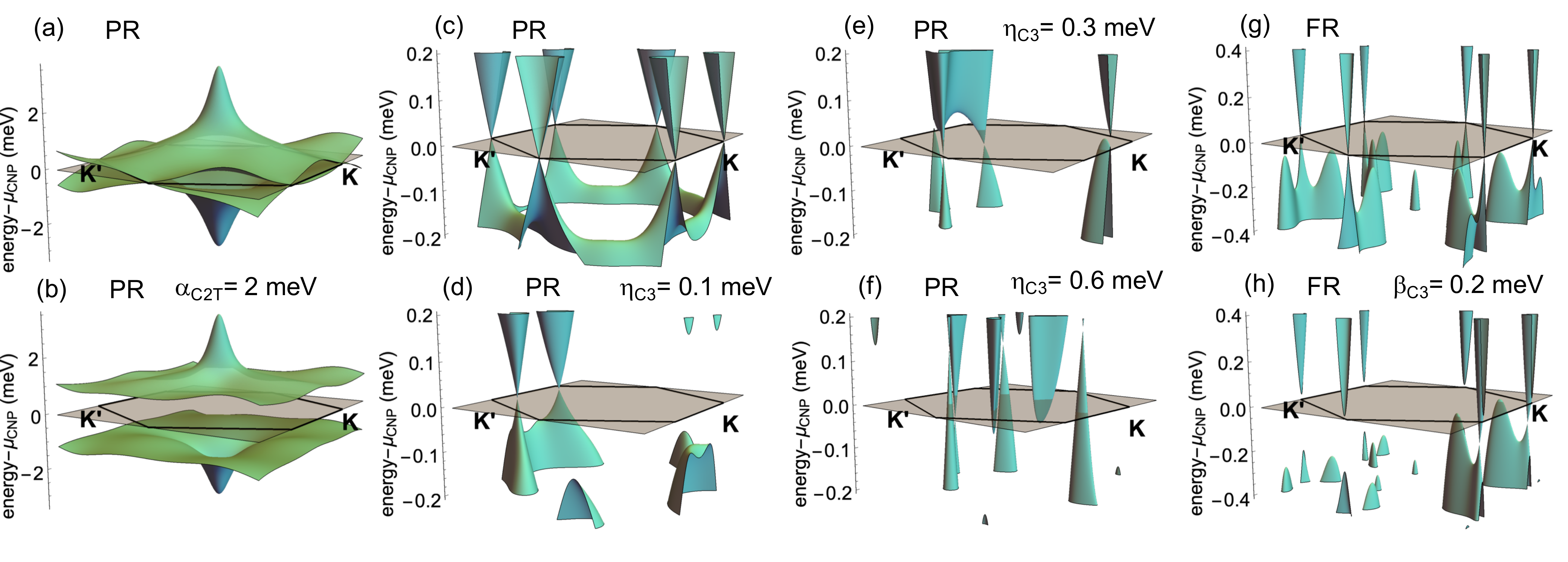}
\caption{{\bf Three dimensional band structure in correlated and non-correlated states.} (a)-(b) Flat bands of the partially relaxed (PR) model  in the (a) non-correlated state, and (b) the C$_2$T symmetry breaking state. The non-correlated state has Dirac points protected by C$_2$T symmetry at the K and K' points as required by the C$_3$ symmetry. When the C$_2$T symmetry is broken, a gap is open at the Dirac points while, in the particular state under study, the region around $\Gamma$ is weakly affected. (c)-(f) Zoom of the flat band structure for the PR model and different values of $\eta_{\rm C_3}$. In (c), $\eta_{\rm C_3}=0$ and the Dirac points are placed at K and K'. 
If $\eta_{\rm C_3}$ is small, as in (d), the Dirac points move in momentum but stay at $\mu_{\rm CNP}$. When its magnitude increases, as in (e) and (f), the bands at small energies are modified displacing the Dirac points away from the chemical potential of the undoped system and creating small Fermi pockets at CNP. Hence, nematicity produces a Lifshiftz transition turning a semi-metal into a metal. 
(g) and (h) Same as (c) and (e) for the FR model in the nematic state $\beta$. In all the panels a plane is plotted at $\mu_{\rm CNP}$ for reference.}
\label{fig:Fig3} 
\end{figure*}

The optical spectrum of the undoped compounds (black curves in Fig.~\ref{fig:Fig1}(d) and (e)) is characterized by the absence of a Drude peak at zero frequency, but with inter-band transitions contributing from zero energy as expected from the presence of Dirac points at $\mu_{\rm CNP}$ \cite{MoonPRB2013,TabertPRB2013,StauberNJP2013}. 
At small frequencies the optical conductivity is controlled by the transitions $\gamma_1$ between the two flat bands and  is finite up to a frequency of the order of the flat bands bandwidth.  
The inter-band transitions are weighted by matrix elements and the maxima in $\sigma'_{\lambda \lambda} (\omega)$ are not simply related to those in the DOS.
For example, in the PR model the conductivity peaks around $\omega \sim 2$ meV while the DOS sharp peaks at $\omega \sim 0.3$ meV, associated with the presence of van Hove singularities at M and M' (see for instance the black line in Fig.~\ref{fig:Fig2}(b)), only produce a small feature in $\sigma'_{\lambda \lambda} (\omega)$ at $\omega \sim 0.6$ meV.

Beyond the frequencies dominated by the $\gamma_1$ transitions, there is a gap up to the frequencies of the $\gamma_2$ and $\gamma_4$ transitions, in Fig.~\ref{fig:Fig1}(c). For the small twist angles considered here the threshold for these transitions is determined by a region in momentum space close to $\Gamma$.  Interestingly, at this threshold there is a sharp peak for the PR model while a small shoulder is found in the FR model. This distinction arises, again, from differences in the matrix elements. 
A smaller peak at higher frequencies signals the onset of $\gamma_3$ transitions. 

In doped TBG new transitions between the flat and higher energy bands ($\gamma_5$ and $\gamma_6$ in Fig.~\ref{fig:Fig1}(c)) are allowed. 
As $\mu$ is moved away from the Dirac points, a gap opens in the optical spectrum corresponding to transitions $\gamma_1$. Nevertheless, if the chemical potential still lies in the flat bands, the system is metallic and there is a Drude peak at zero frequency  (see Supplementary Information). For fully empty or occupied bands, the optical conductivity is zero up to a frequency given by the $\gamma_5$ or $\gamma_6$ transitions, respectively. 

\subsection*{Results in symmetry breaking states}
Fig.~\ref{fig:Fig2} shows how the optical conductivity at CNP responds to different symmetry breaking states. 

In a correlated order which breaks the C$_2$T symmetry, named $\alpha$ here, 
the flat bands at $\Gamma$ are barely affected but the Dirac points at K and K' are gapped~\cite{ManesPRB2007,PoPRX2018,ChoiNatPhys2019}, see Fig.~\ref{fig:Fig3}~(a) and (b), producing activated behavior in transport measurements, as the one observed in some experiments.
This gap shows up in the optical spectrum of the undoped system, see Fig.~\ref{fig:Fig2}~(a), inducing a reorganization of the spectral weight which resembles the one in the DOS shown in Fig.~\ref{fig:Fig2}~(b).

In contrast, a nematic state does not open a gap in the optical conductivity of undoped TBG~\cite{ManesPRB2007,PoPRX2018,ChoiNatPhys2019}. Optical conductivity is a directional probe which reflects the discrete rotational symmetry of a solid. Hence nematicity can be detected by studying the spectrum along different directions. In the absence of correlations or external symmetry breaking the optical conductivity spectra along the directions related by the symmetry rotations of the crystalline lattice are equal. In TBG, the lattice has C$_6$ symmetry when both valleys are included (C$_3$
in each valley) and the optical conductivities along any two axes rotated by $\pm\pi/3$ are equal. Moreover, for any system with C$_3$ symmetry, the equality $\sigma'_{X_iX_i}=\sigma'_{Y_iY_i}$ is also fulfilled. A nematic state lowers the rotational symmetry and the optical conductivity spectra along the affected directions become different. 

\begin{figure*}
\includegraphics[clip,width=1\textwidth]{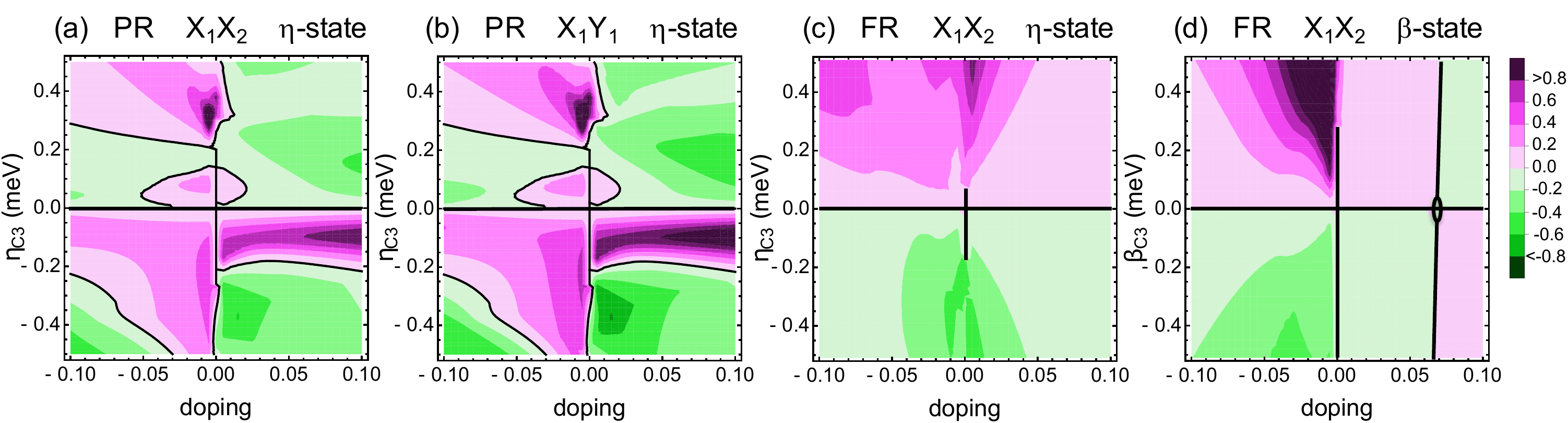}
\caption{ {\bf Anisotropy of the Drude weight.} Colour plots for the (a) $X_1X_2$ and (b) $X_1Y_1$ Drude anisotropy for the PR model in the nematic state $\eta$ as a function of doping and order parameter $\eta_{\rm C_3}$. The $A_iB_j$ anisotropy is defined as the ratio $(D_{A_i}/D_{B_j})-~1$ between the Drude weights $D_{A_i}$ and $D_{B_j}$ along two directions $A_i$ and $B_j$. Doping is defined per spin and valley with respect to the CNP: doping is zero at the CNP and it is 1 and -1 respectively for completely filled and empty bands. (c) and (d) Same as (a) but for the FR model respectively in $\eta$ and $\beta$ nematic states.}
\label{fig:Fig4} 
\end{figure*}

In STM experiments, when the flat bands are partially filled  the $X_1$ direction becomes different to directions $X_2$ and $X_3$ and the DOS is modified~\cite{ChoiNatPhys2019,JiangNat2019,XieNat2019}.
Fig.~\ref{fig:Fig2}~(c) shows the optical conductivity for the nematic order $\eta$ proposed to explain the observed changes in the DOS~\cite{ChoiNatPhys2019}.
As illustrated in  Fig.~\ref{fig:Fig2}~(c) the nematicity is revealed in the spectrum via the inequalities $\sigma'_{X_1X_1}\neq \sigma'_{X_2X_2}=\sigma'_{X_3X_3}$ and $\sigma'_{X_iX_i}\neq \sigma'_{Y_iY_i}$.
The differences show up at frequencies corresponding to the $\gamma_1$, $\gamma_2$, $\gamma_4$ and even $\gamma_3$ transitions, revealing that not only the flat bands but also the higher energy bands are affected by this nematic order. In this state the reorganization of the optical spectrum  in Fig.~\ref{fig:Fig2}~(d) is not easily related to the enhanced peak separation in the DOS shown  in Fig.~\ref{fig:Fig2}~(d).
 The low frequency spectrum depends non-monotonically on the amplitude of the order 
parameter $\eta_{\rm C_3}$
and, as the nematic order enters as an effective hopping term, the spectral weight at low frequencies is not conserved.

Unexpectedly, while the non-interacting state at charge neutrality is a semi-metal, a Drude peak may arise for finite values of $\eta_{\rm C_3}$ evidencing that a semi-metal to 
metal Lifshitz transition takes place. This Lifshitz transition is illustrated in Fig.~\ref{fig:Fig3}~(c) to (f), which show the evolution with $\eta_{\rm C_3}$ of the band structure 
close to $\mu_{\rm CNP}$. In the non-correlated state with $\eta_{\rm C_3}=0$ the Dirac points  are  
placed at K and K'  as required by C$_3$ symmetry (Fig.~\ref{fig:Fig3}~(c)). For small  $\eta_{\rm C_3}$ the Dirac points move in $k$-space from K and K'~\cite{PoPRX2018} but still lie at $\mu_{\rm CNP}$  (Fig.~\ref{fig:Fig3}~(d)). Further modifications of the flat bands with increasing $\eta_{\rm C_3}$  shift the Dirac points away from $\mu_{\rm CNP}$ and generate small hole and electron Fermi pockets (Fig.~\ref{fig:Fig3}~(e)). The value of the order parameter at which the pockets emerge produces changes in the DOS, red curve in Fig.~\ref{fig:Fig2}~(f), compatible with the band widening observed in STM experiments~\cite{ChoiNatPhys2019}.  Additional Fermi pockets appear for larger values of $\eta_{\rm C_3}$, also leading to new band crossings between the lower and upper flat bands (Fig.~\ref{fig:Fig3}~(f)).  These crossings produce pairs of Dirac points with opposite vorticity which drift away in energy and momentum as the nematicity increases. 

The Lifshiftz transition is not specific to the nematic state $\eta$ or to the PR model. As shown in Fig.~\ref{fig:Fig2}~(f), a finite Drude weight $D_{X_1}$ also appears in the optical spectrum of the FR model in a different nematic state $\beta$ and originates in the Fermi pockets in Fig.~\ref{fig:Fig3}~(h). The presence of several maxima in the lower flat band of the FR model (see Fig.~\ref{fig:Fig1}~(b) and Fig.~\ref{fig:Fig3}~(h)) makes it more prone to display Fermi pockets at the CNP even for small order parameters. See Supplementary Information for more details.   

The anisotropy of the Drude weight, more specifically the sign of the $A_iB_j$-anisotropy ratio $(D_{A_i}/D_{B_j})-~1$ between the Drude weights $D_{A_i}$, $D_{B_j}$ of two otherwise equivalent directions $A_i$ and $B_j$, has been used to characterise the nematic state of other compounds, such as iron superconductors~\cite{ChuScience2010, ValenzuelaPRL2010, FernandesPRL2011,BlombergNatComm2013,IshidaPRL2013,BasconesCRP2016}. In Fig.~\ref{fig:Fig4} we plot the $X_1X_2$  and $X_1Y_1$ anisotropy ratios for the PR and FR models in different nematic states as a function of the electronic filling and the nematic order parameter. Black lines denote the values 
with zero anisotropy and delimitate the regions with different signs. The anisotropy vanishes if the order parameter $\eta_{\rm C_3}$ or $\beta_{\rm C_3}$ is zero. The vertical line at the CNP marks a region with no Drude weight. The end of this line indicates the value of the order parameter at which the Lifshiftz transition takes place. The sign of the anisotropy, originating in our approach in the morphology and topology of the Fermi pockets, is similar for $X_1X_2$  and $X_1Y_1$ for a given TBG model and nematic order but it is notably distinct
for different TBG models with the same order or for the same TBG model with different nematic orders: for example, the sign of the $X_1X_2$ anisotropy of the FR model in state $\eta$ is given by the sign of the  order parameter $\eta_{\rm C_3}$ while the PR model anisotropy in the same state  or the FR model anisotropy in state $\beta$ display several sign changes.

\section*{DISCUSSION}

Our results show that the optical conductivity can distinguish different symmetry states and reveal their correlated nature. 
In particular, both the reduction of the rotational symmetry, detected in STM experiments, and the presence of a gap at the CNP, observed in transport measurements, can be directly seen in the optical conductivity spectrum. 
We have shown that the lack of symmetry may be observable not only at low frequencies but also at higher ones. We have restricted our discussion of the transitions to higher energy bands around $\Gamma$ near the threshold for $\gamma_2$ or $\gamma_4$ transitions as they are expected to be more easily identifiable.
At these frequencies the nematic order could be detected. Nevertheless, at higher frequencies the transitions around $K$ or $M$ involving flat and higher energy bands would also be modified by both C$_3$ and C$_2$T symmetry breakings. If these transitions are identified~\cite{HespArXiv2019}, they could be used to study the correlated state. 
Some of the signatures produced by strain or alignment with the substrate would be similar to the ones discussed here, even though the lattice symmetry breaking effects should enter in the model differently. However, the lattice effects on the spectrum should not vary significantly with doping or temperature, allowing the distinction between a lattice or electronic origin.

In the absence of correlations the chemical potential shifts rigidly and the peaks in the spectrum associated to these transitions do not shift in frequency.  However if, as observed in STM experiments, the signatures of the correlated state, namely the breaking of the C$_3$ symmetry or other modifications of the band structure, disappear when the flat bands are empty or full, a frequency shift is expected in the peaks of the spectrum which originate in transitions between the bands  affected by the correlations. Doping away from the charge neutrality point, the transitions $\gamma_1$ between the flat bands become progressively forbidden but the information can still be obtained from the evolution with doping of transitions $\gamma_2$-$\gamma_4$. These transitions  may be  easier to detect experimentally than the $\gamma_1$, even in undoped systems,  as they appear at significantly higher frequencies.  Detection of a shift  with doping of the frequency of the transitions $\gamma_3$, which do not involve the flat bands, would reveal the involvement of the higher energy bands in the correlated state, at present under debate. In general, the study of the different transitions could allow a momentum selective analysis. For example, the transitions $\gamma_2$-$\gamma_4$ around $\Gamma$ dominate the spectrum close to the threshold while those around K or M appear at higher frequencies. On the other hand, the frequency of the $\gamma_1$ transitions,  whose contribution is very sensitive to the electronic filling, is largest at $\Gamma$.

The nematic order can be also detected by looking at the Drude weight anisotropy in optical or transport experiments.  However, unlike for other materials, like iron superconductors~\cite{ChuScience2010, ValenzuelaPRL2010, FernandesPRL2011,BlombergNatComm2013,IshidaPRL2013,BasconesCRP2016}, the anisotropies' sign displays a complicated dependence on both the doping and the order parameter. Such a  sensitivity to details suggests that a strong sample variability could be found in the sign of the anisotropy of TBG.

Our finding of Lifshitz transitions induced by nematicity impacts not only the optical conductivity but many other experimental measurements, including transport, quantum oscillations or STM. The details of the Lifshitz transition would vary in different samples as the degree of relaxation of the lattice significantly affects the shape of the flat bands and hence how they are affected by the nematic order, making the relaxed lattices more sensitive to it. In general, Fermi pockets emerge already for magnitudes of the order parameter whose effect in the flat band structure is small or compatible with the reorganization of the DOS observed experimentally.

The gap observed at charge neutrality in transport experiments could be due to the breaking of C$_2$T symmetry~\cite{XiePRL2020}, discussed here, but also to other electronic orders such as intervalley coherent states or valley/spin polarized orders~\cite{BultinikArXiv2019,LiuArXiv2019,ZhangArXiv2020,CeaArXiv2020}. However, note that different theoretical predictions do not agree on whether the polarized states at CNP are insulating or metallic ~\cite{LiuArXiv2019,ZhangArXiv2020,CeaArXiv2020}. Within the quasiparticle description adopted here, the contribution of different spins and valleys to the spectrum  add. As the spin and valley are not coupled by the hopping, if the gap at CNP is due to valley or  spin polarized order, the $\gamma_1$ transitions should be suppressed for the spin/valley flavor whose bands become empty (filled). Concomitantly, the $\gamma_5$ ($\gamma_6$) peaks will emerge. Therefore, the temperature dependent spectral weight reorganization could help distinguish whether the gap originates in spin/valley polarization versus the breaking of C$_2$T symmetry or intervalley coherence. 

Finally, very recent experiments have shown that for certain fillings the population of the flat bands occurs through a sequence of phase transitions at which a single spin/valley flavor takes all the carriers and becomes completely empty/filled~\cite{ZondinerArXiv2019,WongArXiv2019}. Such phase transitions should also be detectable via optical conductivity as they would lead to a redistribution of the total spectral weight as a function of doping and temperature.

 \section*{METHODS}
\subsection*{Tight Binding models}
We use a 10-band tight binding model for each valley and spin previously proposed to satisfy all the symmetries and the fragile topology of the continuum model of TBG~\cite{PoPRB2019}, and later confirmed by a projection from an ab-initio $k \cdot p$ perturbation model which includes the lattice relaxation~\cite{CarrPRR2019}. Different choices for the parameters reproduce the band structure corresponding to different angles and degrees of lattice relaxation. The model is based on effective moir\'e orbitals, not on carbon atomic orbitals. The ten effective moir\'e orbitals are located at the three different lattices formed by the symmetry points of the TBG (see Fig.~\ref{fig:Fig1}(c)): 3 $p$ orbitals ($p_{zT}$, $p_{+T}$ and $p_{-T}$) at the triangular lattice determined by the AA regions,  4 $p$ orbitals ($p^A_{+H}$, $p^A_{-H}$, $p^B_{+H}$ and $p^B_{-H}$) at the hexagonal lattice of the the AB and BA regions with a $p_{+}$ and a $p_{-}$ at each of the two $A$ and $B$ lattice sites, and three $s$ orbitals, $s_{K1}$, $s_{K2}$ and $s_{K3}$, at the kagome lattice formed by the SP points, with a single orbital at each of the three sites of the kagome lattice. The flat bands, namely the two bands close to zero energy in Fig.~\ref{fig:Fig1} are mostly composed by the p$_{+T}$ and p$_{-T}$ orbitals having, nevertheless, $p_{zT}$ or kagome $s_{\kappa 1}$, $s_{\kappa 2}$ and $s_{\kappa 3}$ character at the $\Gamma$ point. 

The fully relaxed (FR)  model with $\theta_{\rm FR}\sim  0.9^o$ is derived from an ab-initio $k \cdot p$ approximation which includes the effect of in-plane and out of plane relaxation~\cite{CarrPRR2019}. The tight-binding for the  $\theta_{\rm PR}\sim 1.05^o$ partially relaxed (PR) TBG is obtained from the continuum model and only vertical corrugation is introduced through a $15 \%$ suppression of the interlayer hopping between carbon atoms in the same sublattice~\cite{PoPRB2019,KoshinoPRX2018}.

\subsection*{Symmetry breaking states}
The three orders $\alpha$, $\eta$ and $\beta$ are introduced phenomenologically at the mean field level, but the orders $\eta$ and $\alpha$ were obtained self-consistently from a microscopic model in Ref.~\cite{ChoiNatPhys2019}. To modify the energy of the flat bands the three orders involve the p$_{+T}$ and p$_{-T}$ orbitals. The C$_2$T symmetry breaking  $\alpha$ order breaks the degeneracy between $p_{+T}$ and $p_{-T}$ with an onsite diagonal term.  This makes the two atomic sublattices in each layer inequivalent. The nematic states $\eta$ and $\beta$ arise from bond orders:  $\eta$ breaks the C$_3$ symmetry of the interorbital hopping between $p_{+T}$ and $p_{-T}$ while in  $\beta$ the hopping amplitude between the $p_{+T}$ and $p_{-T}$ orbitals and $s_{\kappa 1}$ differs to the corresponding hopping to $s_{\kappa 2}$  and $s_{\kappa 3}$. In other words, in the non-nematic states the hopping amplitudes in the directions joining the nearest neighbor hexagon centers ($Y_1$, $Y_2$ and $Y_3$ directions, see Fig.~\ref{fig:Fig1} (f)) are equal. In these nematic states, the hopping amplitude in the $Y_1$ direction is different from the other two, making this bond inequivalent to the other two~\cite{ChoiNatPhys2019}. See Supplementary Information for further details. 

\subsection*{Optical Conductivity}
In the optical conductivity calculation we assume a band picture and neglect possible excitonic effects. In a strongly correlated metallic state with both quasiparticle and Hubbard bands, this kind of description will approximate the behavior of the quasiparticle spectrum 
while Hubbard bands would give further contributions~\cite{BasovRevModPhys2011}.
In the derivation of the explicit expression, Supplementary Equation 20,  for multiorbital systems~\cite{ValenzuelaPRB2013} the coupling to the electromagnetic field is introduced via a Peierls substitution. Within the used approximation the matrix elements for the optical conductivity are written as derivatives of the orbital-dependent tight binding terms. The 10-band model used here has several atoms per unit cell and applicability of the expressions in Ref.~\cite{ValenzuelaPRB2013} requires writing the Hamiltonian such that the specific distances between the atoms are accounted for, as the phase acquired in the Peierls substitution is sensitive to these distances and not to unit cell labels~\cite{NguyenPRB2016}, see Supplementary Information. Provided the valley and spin degeneracies are not broken,  in the used approximation, the optical conductivities of both valleys and spins are equal. We therefore focus on the optical conductivity of a single valley and spin. A small  broadening  0.04  meV is used in the calculation of $\sigma_{\lambda \lambda}(\omega)$ except in Fig.~\ref{fig:Fig2}(e) and (f) where it is 0.06  meV. The Drude weight is calculated directly from the hamiltonian and the current matrix elements without any broadening, see Supplementary Information.

\section*{DATA AVAILABILITY}
All relevant data are available from the authors upon reasonable request.

\section*{ACKNOWLEDGMENTS}
We thank conversations with E. Andrei, D. Basov, P. Jarillo-Herrero, A. H. MacDonald, H. Ch. Po, P. San Jos\'e, O. Vafek, and A. Vishwanath. Funding from Ministerio de Ciencia, Innovaci\'on y Universidades via Grants FIS2015-64654-P and PGC2018-097018-B-I00.  

\section*{COMPETING INTERESTS}
The authors declare no competing interests.

\section*{AUTHOR CONTRIBUTIONS}
E.B. conceived the project. Both M.J.C. and E.B. designed and performed the calculations, analysed and discussed the results, and wrote the manuscript.

\beginsupplement
\begin{widetext}
\section{SUPPLEMENTARY INFORMATION}

\subsection*{Tight-binding model}
We describe the non-interacting band-structure with a 10-band tight-binding model per valley based on effective moir\'e orbitals. The ten effective moir\'e orbitals are located at the three different lattices formed by the symmetry points of the TBG (see Fig.~1(f)): 
 3 $p$ orbitals ($p_{zT}$, $p_{+T}$ and $p_{-T}$) at the triangular lattice determined by the AA regions,  4 $p$ orbitals ($p^A_{+H}$, $p^A_{-H}$, $p^B_{+H}$ and $p^B_{-H}$) at the hexagonal lattice formed by the AB and BA regions with a $p_{+}$ and a $p_{-}$ at each of the two inequivalent $A$ and $B$ lattice sites, and three $s$ orbitals, $s_{K1}$, $s_{K2}$ and $s_{K3}$, at the kagome lattice formed by the SP points, with a single orbital at each of the three sites of the kagome lattice. 

This tight-binding was introduced in Ref.~\cite{PoPRB2019} where parameters were chosen to fit the bands obtained from the continuum theory for TBG with a twist angle of $\theta = 1.05^o$~\cite{KoshinoPRX2018}. These parameters correspond to what we call PR (partially relaxed) model in the main text. The same tight-binding was used in Ref.~\cite{CarrPRR2019} where it was fitted to the band-structure resulting from an ab-initio $k \cdot p$ continuum model which account for the lattice relaxation for a  twist angle of $\theta = 0.9^o$. We use these parameters for our FR (fully relaxed) model. The parameters for the PR and FR models are listed in Table~\ref{table:param} and the corresponding band structure is plotted in Fig.~\ref{fig:FigS1}. The in-plane relaxation strongly reduces the approximate particle-hole symmetry of the continuum model and results in a more complicated flat bands dispersion. In both cases, the considered angles are below the corresponding magic angles ($1.08^o$ for PR and $1^o$ for FR). 

Here we write down the $10\times10$ hamiltonian in the symmetric formulation required to calculate the optical conductivity. It is useful to define
\begin{equation}
\phi_{lm}= e^{-i {\bf k} \cdot (l {\bf a}_1+ m {\bf a}_2)}
\end{equation}
with ${\bf a}_1=a \hat y$ and ${\bf a}_2=a \left(\frac{\sqrt{3}}{2} \hat x -\frac{1}{2} \hat y \right)$ the lattice vectors, $l,m$ fractional numbers and $a$ the moir\'e lattice constant. $\phi_{lm}$ condenses the information about the hopping directions. The tight binding hamiltonian is written $H=\Sigma_{\bf k}\Phi^\dagger h_{10 \times 10}({\bf k})\Phi $ with
\begin{equation}
\Phi=(p_{zT},p_{+T},p_{-T},s_{\kappa 1}, s_{\kappa 2},s_{\kappa 3},p^A_{+H},p^A_{-H},p^B_{+H},p^B_{-H}) .
\end{equation}

For convenience, in order to write $h_{10 \times 10}({\bf k})$ we differentiate the six orbitals at the triangular and kagome lattices from the 4 orbitals at the hexagonal lattice. In the following we do not write explicitly the dependence on ${\bf k}$. 
\begin{equation}
h_{10 \times 10}=\quad
\begin{pmatrix}
H_{6\times 6} & H_{6 \times 4} \\
H^\dagger_{6 \times 4} & H_{4\times 4} 
\end{pmatrix}
\quad .
\end{equation}

Here 
\begin{equation}
H_{6\times 6}=\quad
\begin{pmatrix}
H_{p_z}+\mu_{p_z} & C^\dagger_{p_{\pm}p_z} & 0  \\
C_{p_{\pm}p_z} & H_{p_\pm}+\mu_{p_\pm}\hat{\bf1}_{2\times2} & C^\dagger_{\kappa p_{\pm}} \\
0 & C_{\kappa p_{\pm}} & H_{\kappa}+\mu_{\kappa}\hat{\bf1}_{3\times3} 
\end{pmatrix}
\quad ,
\end{equation}
$H_{pz}$ describes the $p_{zT}$ intraorbital term
\begin{equation}
H_{p_z}=t_{p_{z}} \left( \phi_{01}+\phi_{11} + \phi_{10} + h.c. \right) .
\end{equation}
$H_{\pm}$ gives the intra and interorbital hopping between the $p_{\pm T}$ orbitals 
\begin{equation}
H_{p_{\pm}}=\quad\begin{pmatrix}
t_{p_{\pm}}\left( \phi_{01}+\phi_{11} + \phi_{10} + h.c. \right) & C_{p_\pm p_\pm}^\dagger \\
C_{p_\pm p_\pm} & t_{p_{\pm}}\left( \phi_{01}+\phi_{11} + \phi_{10} + h.c. \right) 
\end{pmatrix}
\quad
\end{equation}
with 
\begin{equation}
C_{p_\pm p_\pm} = t^+_{p_\pm p_\pm} \left( \phi_{01}+\phi_{\bar 1 \bar 1}\Omega + \phi_{10}\Omega^* \right)+ t^-_{p_\pm p_\pm} \left( \phi_{0 \bar 1}+\phi_{ 11}\Omega + \phi_{\bar 10}\Omega^* \right)
\end{equation}
and $C_{p_{\pm}p_z}$ the interorbital hopping between $p_{zT}$ and $p_{\pm T}$:
\begin{equation}
C_{p_{\pm}p_z}=i t^+_{p_{\pm}p_z} \quad\begin{pmatrix}
\phi_{01}+\phi_{\bar 1 \bar 1} \Omega + \phi_{10}\Omega^* \\
-\left(\phi_{0\bar 1}+\phi_{11} \Omega^*+ \phi_{\bar 10} \Omega \right) 
\end{pmatrix}
\quad
- i t^-_{p_{\pm}p_z} \quad\begin{pmatrix}
\phi_{0\bar 1}+\phi_{11} \Omega + \phi_{\bar 10}\Omega^* \\
-\left(\phi_{01}+\phi_{\bar 1\bar 1} \Omega^*+ \phi_{10} \Omega \right) 
\end{pmatrix}
\quad.
\end{equation}
$H_\kappa$ includes the hopping between the orbitals in the kagome lattice 
\begin{equation}
H_\kappa=t_\kappa \quad
\begin{pmatrix}
0 & \phi_{-\frac{1}{2}0}+ \phi_{\frac{1}{2}0}  & \phi_{-\frac{1}{2}-\frac{1}{2}}+\phi_{\frac{1}{2}\frac{1}{2}} \\
\phi_{-\frac{1}{2}0}+ \phi_{\frac{1}{2}0} & 0 & \phi_{0-\frac{1}{2}}+ \phi_{0\frac{1}{2}} \\
 \phi_{-\frac{1}{2}-\frac{1}{2}}+\phi_{\frac{1}{2}\frac{1}{2}}  & \phi_{0-\frac{1}{2}}+ \phi_{0\frac{1}{2}} & 0
\end{pmatrix}
\quad
+t'_\kappa \quad
\begin{pmatrix}
0 & \phi_{\frac{1}{2}1}+ \phi_{-\frac{1}{2}-1}  & \phi_{\frac{1}{2}-\frac{1}{2}}+\phi_{-\frac{1}{2}\frac{1}{2}} \\
\phi_{\frac{1}{2}1}+ \phi_{-\frac{1}{2}-1} & 0 & \phi_{1\frac{1}{2}}+ \phi_{-1-\frac{1}{2}} \\
 \phi_{\frac{1}{2}-\frac{1}{2}}+\phi_{-\frac{1}{2}\frac{1}{2}}  & \phi_{1\frac{1}{2}}+ \phi_{-1-\frac{1}{2}} & 0
\end{pmatrix}
\quad,
\end{equation}
and $C_{\kappa p_\pm}$ between $s_{\kappa (1,2,3)}$ and $p_{\pm T}$ terms are:
\begin{equation}
C_{\kappa p_\pm}=t_{\kappa p_\pm}^{+}
 \quad
\begin{pmatrix}
\phi_{0\frac{1}{2}} & \phi_{0-\frac{1}{2}} \\
\phi_{-\frac{1}{2}-\frac{1}{2}}\Omega^* &  \phi_{\frac{1}{2}\frac{1}{2}}\Omega \\
 \phi_{\frac{1}{2}0} \Omega & \phi_{-\frac{1}{2}0} \Omega^*  
\end{pmatrix}
\quad
-t_{\kappa p_\pm}^{-}
 \quad
\begin{pmatrix}
\phi_{0-\frac{1}{2}} &  \phi_{0\frac{1}{2}}  \\
\phi_{\frac{1}{2}\frac{1}{2}}\Omega^* &  \phi_{-\frac{1}{2}-\frac{1}{2}}\Omega \\
 \phi_{-\frac{1}{2}0}\Omega & \phi_{\frac{1}{2}0}\Omega^*  
\end{pmatrix}
\quad.
\end{equation}
The onsite energies are written in terms of the hopping parameters: $\mu_{p_z} =-6 t_{p_z} +\delta_{p_z}$, $\mu_{p_\pm}=  3 t_{p_\pm} + \delta_{p_\pm}$, and $\mu_\kappa=-4 (t_\kappa+t'_\kappa)+ \delta_\kappa$. 

\begin{figure*}
\includegraphics[clip,width=0.45\textwidth]{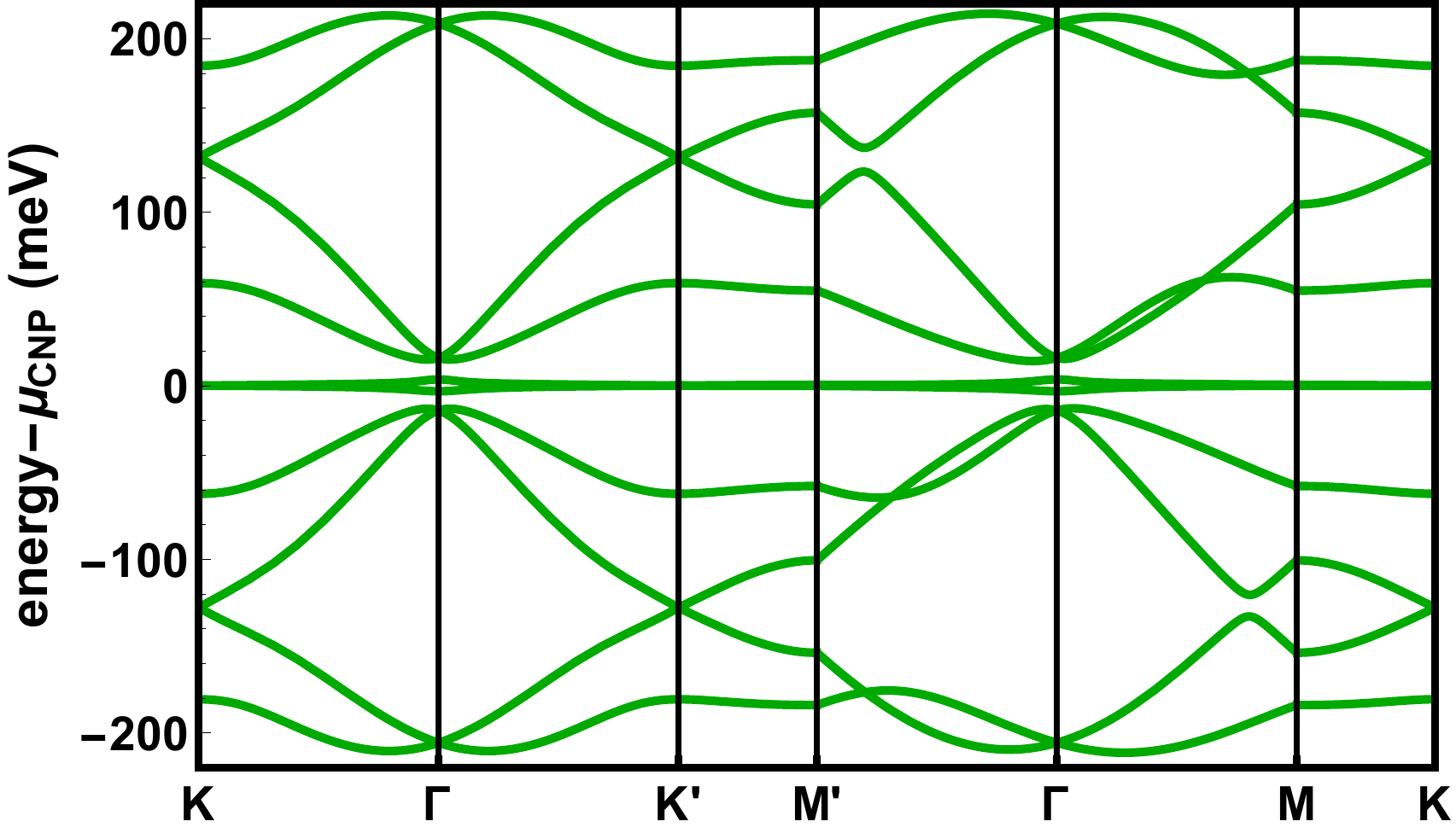}
\includegraphics[clip,width=0.45\textwidth]{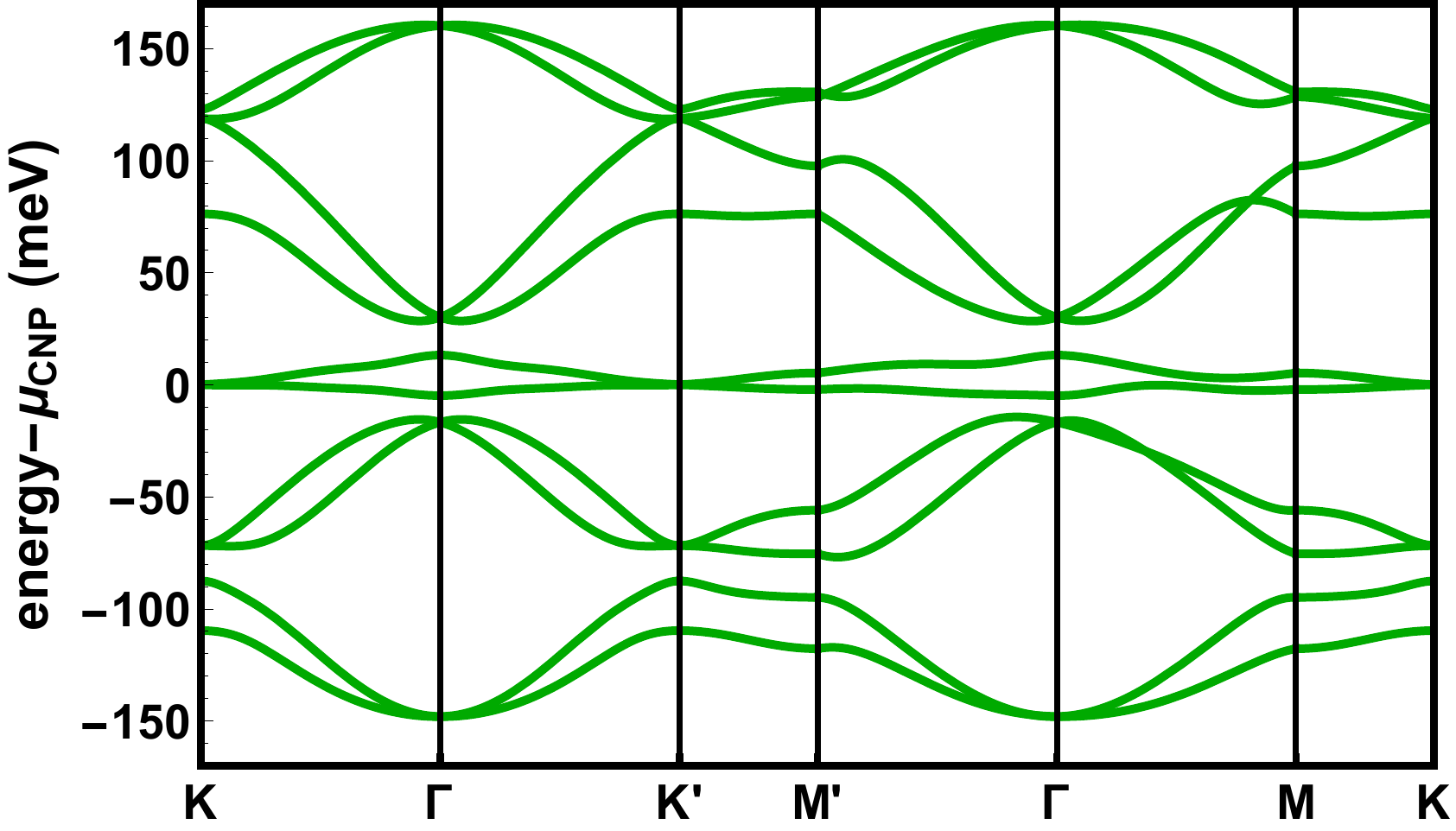}
\caption{Full band structure of the considered models in the uncorrelated state. (a) Bands corresponding to the partially relaxed (PR) model and (b) to the fully relaxed (FR) model. A zoom of the flat bands is shown in Fig.~1 in the main text.} 
\label{fig:FigS1} 
\end{figure*}

The hopping  between these six orbitals and the four orbitals in the hexagonal lattice is given by $H_{6 \times 4}$
\begin{equation}
H_{6 \times 4}=\left(h_{p^+}^{(A)} \,\,\,\,\, h_{p^-}^{(A)}  \,\,\,\,\,  h_{p^+}^{(B)}  \,\,\,\,\, h_{p^-}^{(B)} \right) \, .
\end{equation}
Here
\begin{equation}
h_{p^+}^{(A)}=\quad
\begin{pmatrix}
-\left(\Omega \phi_{-\frac{1}{3}-\frac{2}{3}}+\Omega^*  \phi_{\frac{2}{3}\frac{1}{3}}+ \phi_{-\frac{1}{3}\frac{1}{3}}\right) \zeta^* a \\
\left(\Omega^* \phi_{-\frac{1}{3}-\frac{2}{3}}+\Omega  \phi_{\frac{2}{3}\frac{1}{3}}+ \phi_{-\frac{1}{3}\frac{1}{3}}\right) \zeta b \\
 \left( \phi_{-\frac{1}{3}-\frac{2}{3}}+  \phi_{\frac{2}{3}\frac{1}{3}}+ \phi_{-\frac{1}{3}\frac{1}{3}}\right) c \\
 -i \phi_{-\frac{1}{3}-\frac{1}{6}} d \\
 -i \Omega \phi_{\frac{1}{6}-\frac{1}{6}} d \\
 -i \Omega^*\phi_{\frac{1}{6}\frac{1}{3}} d
\end{pmatrix}
\quad,
\end{equation}
\begin{equation}
h_{p^-}^{(A)}=\quad
\begin{pmatrix}
-\left(\phi_{-\frac{1}{3}-\frac{2}{3}}+\Omega^*  \phi_{\frac{2}{3}\frac{1}{3}}+ \Omega \phi_{-\frac{1}{3}\frac{1}{3}}\right) \zeta^* a \\
 \left( \phi_{-\frac{1}{3}-\frac{2}{3}}+  \phi_{\frac{2}{3}\frac{1}{3}}+ \phi_{-\frac{1}{3}\frac{1}{3}}\right) c \\
\left( \phi_{-\frac{1}{3}-\frac{2}{3}}+\Omega  \phi_{\frac{2}{3}\frac{1}{3}}+ \Omega^*\phi_{-\frac{1}{3}\frac{1}{3}}\right) \zeta b \\
 -i \phi_{-\frac{1}{3}-\frac{1}{6}} d \\
 -i \Omega^* \phi_{\frac{1}{6}-\frac{1}{6}} d \\
 -i \Omega \phi_{\frac{1}{6}\frac{1}{3}} d
\end{pmatrix}
\quad , 
\end{equation}
\begin{equation}
h_{p^+}^{(B)}=\quad
\begin{pmatrix}
-\left(\Omega \phi_{-\frac{2}{3}-\frac{1}{3}}+\Omega^*  \phi_{\frac{1}{3}-\frac{1}{3}}+ \phi_{\frac{1}{3}\frac{2}{3}}\right) \zeta a \\
\left(\Omega^* \phi_{-\frac{2}{3}-\frac{1}{3}}+\Omega  \phi_{\frac{1}{3}-\frac{1}{3}}+ \phi_{-\frac{1}{3}\frac{2}{3}}\right) \zeta^* b \\
 \left( \phi_{-\frac{2}{3}-\frac{1}{3}}+  \phi_{\frac{1}{3}-\frac{1}{3}}+ \phi_{\frac{1}{3}\frac{2}{3}}\right) c \\
 i \phi_{\frac{1}{3}\frac{1}{6}} d \\
 i \Omega \phi_{-\frac{1}{6}\frac{1}{6}} d \\
 i \Omega^* \phi_{-\frac{1}{6}-\frac{1}{3}} d
\end{pmatrix}
\quad ,
\end{equation}
and
\begin{equation}
h_{p^-}^{(B)}=\quad
\begin{pmatrix}
-\left(\Omega \phi_{-\frac{2}{3}-\frac{1}{3}}+  \phi_{\frac{1}{3}-\frac{1}{3}}+\Omega^* \phi_{\frac{1}{3}\frac{2}{3}}\right) \zeta a \\
 \left( \phi_{-\frac{2}{3}-\frac{1}{3}}+  \phi_{\frac{1}{3}-\frac{1}{3}}+ \phi_{\frac{1}{3}\frac{2}{3}}\right) c \\
\left(\Omega^* \phi_{-\frac{2}{3}-\frac{1}{3}}+\Omega  \phi_{\frac{1}{3}-\frac{1}{3}}+ \phi_{\frac{1}{3}\frac{2}{3}}\right) \zeta^* b \\
 i \phi_{\frac{1}{3}\frac{1}{6}} d \\
 i \Omega^* \phi_{-\frac{1}{6}\frac{1}{6}} d \\
 i \Omega \phi_{-\frac{1}{6}-\frac{1}{3}} d
\end{pmatrix}
\quad ,
\end{equation}
with $\zeta=e^{i 2\pi/6}$ and $\Omega=\zeta^2$.
Finally, the terms within the hexagonal lattice are
\begin{equation}
H_{4\times 4}=t_{h} \quad
\begin{pmatrix}
0 & e^{i\phi_{h}} \left(\phi_{-\frac{1}{3}\frac{1}{3}}+\phi_{\frac{2}{3}\frac{1}{3}}+\phi_{-\frac{1}{3}-\frac{2}{3}}\right) \\
e^{-i\phi_{h}} \left(\phi_{\frac{1}{3}-\frac{1}{3}}+\phi_{-\frac{2}{3}-\frac{1}{3}}+\phi_{\frac{1}{3}\frac{2}{3}}\right) & 0 
\end{pmatrix}
\quad \otimes \hat {\bf 1}_{2\times2} .
\end{equation}

\begin{table}
\caption{Values of the parameters used for the two models considered in the main text. We use the notation in Ref.~\cite{PoPRB2019}.}
\label{table:param}
\begin{tabular}{|c|c|c|}
\hline
Parameter &  PR model~\cite{PoPRB2019} (meV) & FR model~\cite{CarrPRR2019} (meV) \\ \hline
$\delta_{p_z}$ & -3.25 & -12.1 \\ \hline
$\delta_{p_\pm}$ & 0 & 6.467  \\ \hline
$\delta_{\kappa}$ &  3.575 & 5.933  \\ \hline
$t_{p_z}$ & 0. & -3.661  \\ \hline
$t_{p_\pm}$ & 0.0975 & -0.205  \\ \hline
$t_{p_\pm p_\pm}^{+}$ & 0 & -2.26  \\ \hline
$t_{p_\pm p_\pm}^{-}$ & 0.13 & -1.661  \\ \hline
$t_{\kappa}$ & 0 & 8.989 \\ \hline
$t'_{\kappa}$ & 0 & -6.634 \\ \hline
$t_{p_\pm p_z}^{+}$ & 0.52 & 3.831  \\ \hline
$t_{p_\pm p_z}^{-}$ & 0 & 1.149  \\ \hline
$t_{\kappa p_\pm}^{+}$ & 0.52 & -7.956  \\ \hline
$t_{\kappa p_\pm}^{-}$ & -0.52 & -5.678  \\ \hline
$t_{h}e^{i \phi_h}$ & 32.5 i & i \\ \hline
$a$ & 14.3  & -24.69 \\ \hline
$b$ & 4.29  & -22.89 \\ \hline
$c$ & 4.29  & -10.51 \\ \hline
$d$ & 74.49 &  36.673 \\ \hline
\end{tabular}
\end{table}

\subsection*{Symmetry breaking states}

To study the optical response to symmetry breaking correlated states we introduce phenomenologically three different terms which break any symmetry of the Hamiltonian  of a given valley:

\begin{equation}
H_{\alpha}=\alpha_{\rm C_2T}\Sigma_{\bf k}( p^\dagger_{+,T,{\bf k}}p_{+,T,{\bf k}}-p^\dagger_{-,T,{\bf k}}p_{-,T,{\bf k}})
\end{equation}
\begin{equation}
H_{\eta}=\eta_{C_3}\Sigma_{\bf k}(\phi_{01} +\phi_{10} +\phi_{-1-1}) p^\dagger_{+,T,{\bf k}}p_{-,T,{\bf k}}+h.c
\end{equation}
\\

\begin{equation}
H_{\beta}=
\beta_{C_3}\Sigma_{\bf k}\left(s^\dagger_{\kappa_1,{\bf k}}  \,s^\dagger_{\kappa_2,{\bf k}}  \,  s^\dagger_{\kappa_3,{\bf k}}  \right) 
\begin{pmatrix}
-\phi_{0-\frac{1}{2}} &  -\phi_{0\frac{1}{2}}  \\
0.5\phi_{\frac{1}{2}\frac{1}{2}}\Omega^* &  0.5\phi_{-\frac{1}{2}-\frac{1}{2}}\Omega \\
0.5 \phi_{-\frac{1}{2}0}\Omega & 0.5\phi_{\frac{1}{2}0}\Omega^*  
\end{pmatrix}
\begin{pmatrix}
 p_{+,T,{\bf k}}  \\
 p_{-,T,{\bf k}} 
\end{pmatrix}
+ h.c.
\end{equation}

$H_\alpha$ breaks the C$_2$T and  the $M_{2y}$ symmetries while $H_\eta$ and $H_\beta$ break the C$_3$ symmetry~\cite{ChoiNatPhys2019}. These terms can be interpreted as the interaction part of a mean-field hamiltonian and are introduced to explore some generic features of the optical conductivity and the modification of the band structure under a symmetry breaking phase transition. Nevertheless $H_\alpha$ and $H_\eta$ were obtained in a self-consistent calculation in Ref.~\cite{ChoiNatPhys2019} using a model based on the PR tight-binding with a Hubbard interaction restricted to the three p-orbitals in the triangular lattice. The state $\eta$, namely with finite $\eta_{C_3}$, was proposed to explain the enhanced peak separation in the density of states observed when the chemical potential lies in the flat bands, as compared to empty or full bands, observed in this work~\cite{ChoiNatPhys2019}. Fig.~S2 shows the modification of the bands  in the ordered states $\eta$ and $\beta$ along several directions in $k$-space, with Fermi pockets at the CNP, as in Fig. 3. These pockets may emerge even for a quite moderate modification of the band structure: an order parameter $\beta_{C_3}$ three times smaller than the one used in  Fig.~S2(b) moves the Dirac points below the CNP in Fig. 3(h). When $\alpha_{C_{2}T}$, $\eta_{C_3}$ and $\beta_{C_3}$ vanish we refer to the system as non-correlated. 

With these terms we aim to represent the symmetry breaking effects associated to an electronic transition. Nevertheless some of the features present in the optical conductivity are generic for a given symmetry breaking independently of whether the symmetry breaking is due to the electronic interactions or a lattice effect, even if the latter should be modelled with different terms in the hamiltonian. The alignment with h-BN has two effects on twisted bilayer graphene: (i) it induces a weak moir\'e potential due to the different lattice constant of h-BN and graphene, and (ii) it breaks the sublattice symmetry (and consequently the C$_2$T symmetry in each valley) due to the different potential exerted by the boron and nitrogen atoms on the two carbon sublattices~\cite{MoonPRB2014}. The second effect opens a gap at the Dirac points, as also found in the $\alpha$ order.  Uniaxial strain in the device would break the $C_3$ symmetry making one of the directions $X_i$ different to the other two and producing an anisotropic response in the optical conductivity while shifting the Dirac points from $K$ and $K'$ as in $\eta$ and $\beta$ states. If the strain is different in the two layers (heterostrain), the Dirac points would not remain degenerate~\cite{HuderPRL2018, BiPRB2019}.

\begin{figure*}
\includegraphics[clip,width=0.90\textwidth]{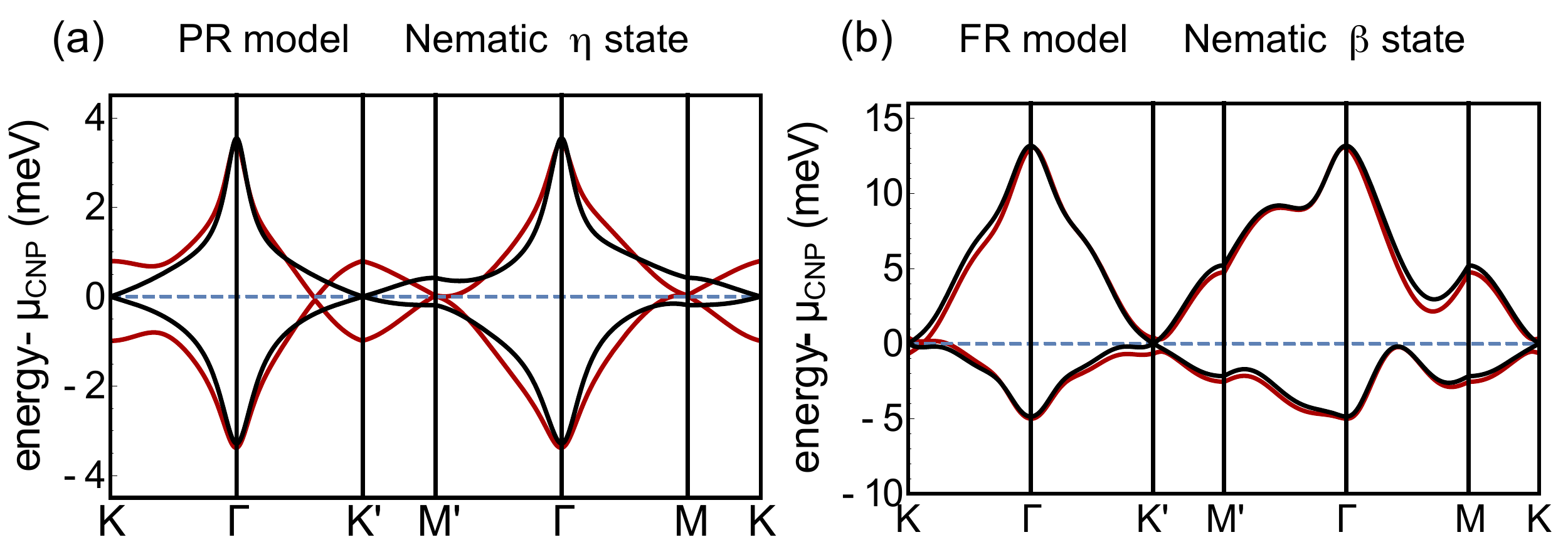}
\caption{{\bf Flat bands in the nematic states.} Red lines in (a) and (b) respectively show the low energy bands corresponding to the partially relaxed (PR) model in nematic state $\eta$ with $\eta_{C_3}=0.3$ meV. (a) and to the fully relaxed (FR) model in state $\beta$ with $\beta_{C_3}=0.6$ meV. (b). For comparison the black lines in both plots display the energy bands for the same models in the non-correlated states. A zoom of the bands in (a) is shown in Fig.~3(e) in the main text. } 
\label{fig:FigS2} 
\end{figure*}

\subsection*{Optical Conductivity}
Here we reproduce the expressions for the conductivity used in the calculations and derived in Ref.~\cite{ValenzuelaPRB2013} for a Hamiltonian $H_{\rm mf}$ bilinear in fermionic operators using Kubo formula and introducing the coupling between the electrons and the electromagnetic field via a  Peierls substitution:
\begin{eqnarray}
&\sigma^\prime_{\lambda \lambda}(\omega)&=D_{\lambda}\delta(\omega)
+\frac{\pi}{V}\sum_{{\bf k}n \neq n'} \frac{\left |j^\lambda_{n'n}({\bf k}) \right|^2}{\epsilon_{n'}({\bf k})-\epsilon_{n}({\bf k})} 
\\&&
\times \theta(\epsilon_{n'}({\bf k})) \theta(-\epsilon_{n}({\bf k})) \delta(\omega-\epsilon_{n'}({\bf k})+\epsilon_{n}({\bf k})), \nonumber
\label{eq:optcondband}
\end{eqnarray}
$\epsilon_{n}(\bf {k})$ and $\epsilon_{n'}(\bf {k})$ are the band energies, $\theta(\epsilon_{n}({\bf k}))$ the Heaviside step function and the 
Drude weight $D_{\lambda}$ can be written
\begin{eqnarray}
&D_{\lambda}&=-\pi \sum_{{\bf k}n}t_{nn}^{\lambda\lambda}({\bf k})\theta(-\epsilon_{n}({\bf k})) 
\nonumber \\&&
-\frac{2\pi}{V}\sum_{{\bf k}n \neq n'} \frac{\left |j^\lambda_{n'n}({\bf k})\right|^2}{\epsilon_{n'}({\bf k})-\epsilon_{n}({\bf k})} \theta(\epsilon_{n'}({\bf k})) \theta(-\epsilon_{n}({\bf k}))  \, ,
\label{eq:drude2}
\end{eqnarray}
with
\begin{eqnarray}
&t_{nn}^{\lambda\lambda}({\bf k})&=\sum_{\mu\nu}\frac{\partial^2 \epsilon_{\mu\nu}({\bf k})}{\partial k_{\lambda}^2}a_{\mu n}^*({\bf k})a_{\nu n}({\bf k}) \, ,
\\
&j_{n'n}^\lambda (\bf {k})&=-\sum_{\mu\nu}\frac{\partial \epsilon_{\mu\nu}({\bf k})}{\partial k_\lambda}a_{\mu n'}^*({\bf k})a_{\nu n}({\bf k}) \, ,
\label{eq:optvertex}
\end{eqnarray}
with $\epsilon_{\mu\nu}$ the elements of the hamiltonian written in the orbital basis 
$H_{\rm mf}=\Sigma_{\bf k,\mu,\nu} \epsilon_{\mu \nu} ({\bf k})c^\dagger_{\mu,{\bf k}} c_{\nu,{\bf k}}=\Sigma_{\bf k, n} \epsilon_{n} ({\bf k})d^\dagger_{n,{\bf k}} d_{n,{\bf k}}$
and $a_{\mu n}({\bf k})$ the rotation matrix between the orbital and the band basis $c_{{\bf{k}}\mu \sigma}^\dagger=\sum_n a_{\mu n}^*({\bf k})d^\dagger_{{\bf {k}}n \sigma}$.

In the present case $c^\dagger_{\mu,{\bf k}}$ and $c_{\nu,{\bf k}}$ are the creation and annihilation operators for the ten orbitals $p_{zT}$, $p_{+T}$, $p_{-T}$, $s_{\kappa 1}$,  $s_{\kappa 2}$, $s_{\kappa 3}$, $p^A_{+H}$, $p^A_{-H}$, $p^B_{+H}$ and $p^B_{-H}$.  $H_{\rm mf}$ includes both the tight-binding $h_{10 \times 10}({\bf k})$ and the mean-field terms $H_\eta$ or $H_\beta$.  $H_\eta$ and $H_\beta$ can be written as hopping terms and couple to the electromagnetic field, contributing to the vertices $t_{nn}^{\lambda\lambda}({\bf k})$ and $j_{n'n}^\lambda (\bf {k})$, while $H_\alpha$ is a local onsite term and its contribution to the vertices vanishes. As shown in Fig.~S3,  in the absence of correlations and with the chemical potential within the flat bands, the low energy interband transitions $\gamma_1$ become progressively suppressed and  a Drude peak emerges as the system is doped from the CNP.  The transitions $\gamma_6$, whose threshold correspond to transitions around $\Gamma$, are easily detected only when the bands are completely filled. At the same doping, the transitions $\gamma_4$ become completely suppressed.  

\begin{figure*}
\includegraphics[clip,width=0.45\textwidth]{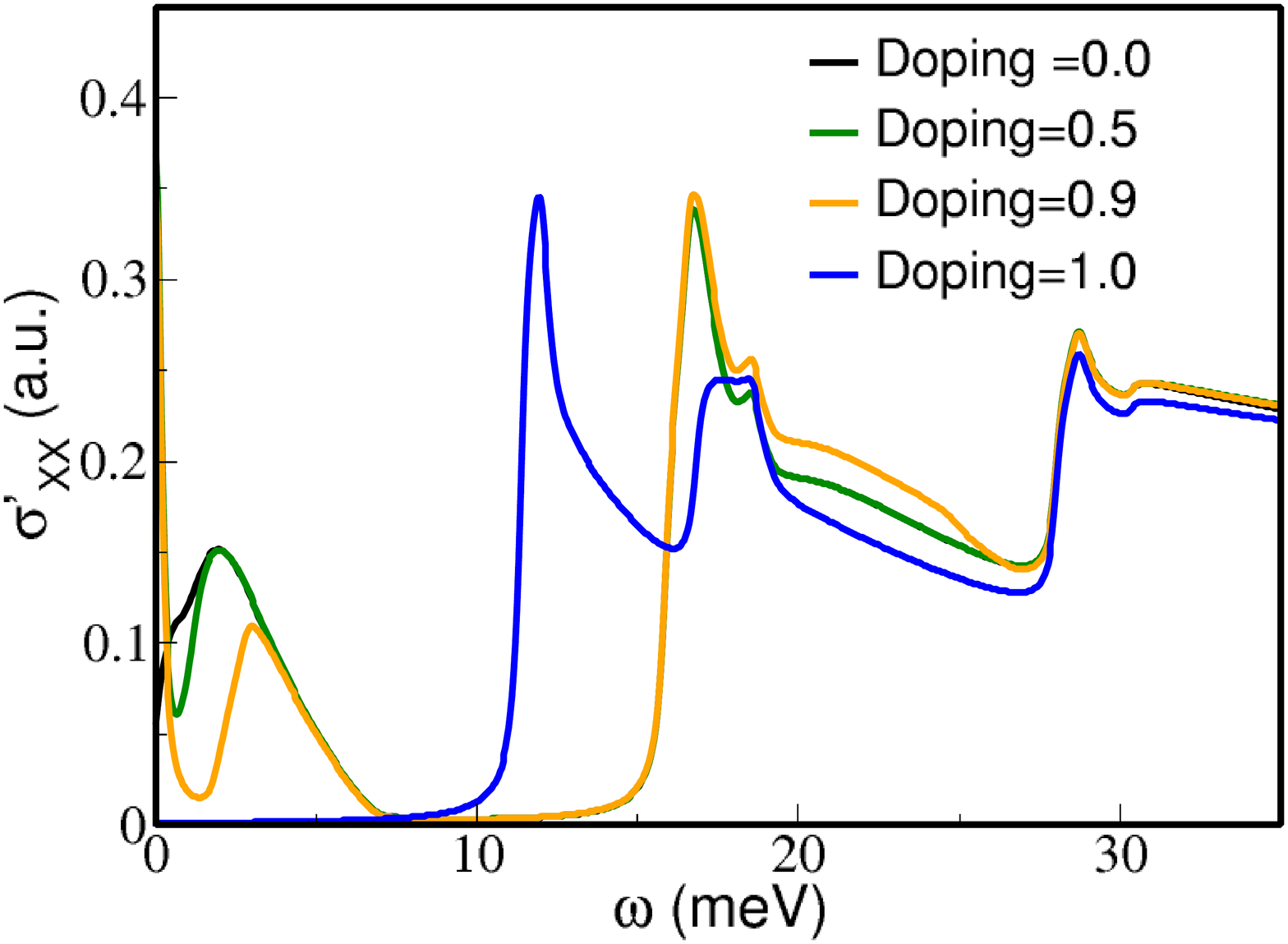}
\caption{{\bf Optical conductivity of the PR model as a function of electronic filling.} Optical conductivity for the non-interacting bands for the partially relaxed model at the CNP (black), with 0.50 (green) and 0.90 (orange) and 1.0 (blue) electrons per spin and valley, see Fig.1 in the main text for other fillings. The broadening used here is $0.2$ meV.} 
\label{fig:FigS3} 
\end{figure*}
\end{widetext}
\bibliography{optcond}

\end{document}